\documentstyle[aps,prb,epsf]{revtex}

\begin{document}

\twocolumn[\hsize\textwidth\columnwidth\hsize\csname
@twocolumnfalse\endcsname

\title{Liquid-liquid equilibrium for monodisperse spherical particles}
\author{E. A. Jagla}
\address{Centro At\'omico Bariloche and Instituto Balseiro, 
Comisi\'on Nacional de Energ\'{\i}a
At\'omica \\(8400) S. C. de Bariloche, Argentina}
\maketitle
 
\begin{abstract}
A system of identical particles interacting through an isotropic
potential that allows for two preferred interparticle distances is numerically 
studied. When the parameters of the interaction potential are adequately chosen, 
the system exhibits coexistence between
two different liquid phases (in addition to the usual liquid-gas coexistence). It is shown that
this coexistence
can occur at equilibrium, namely, in the region where the liquid is thermodynamically stable. 

\end{abstract}
 
\vskip2pc] \narrowtext


\section{Introduction}

Traditionally, liquid-liquid separation for single component systems was not
considered as a real possibility, because of
the rapidly changing liquid structure,
which smoothly varies on temperature and pressure\cite{mcmillan}. This
view is changing. Experimentally, a whole family of network forming fluids,
including some tetrahedrally coordinated materials, is suspected to have
transitions between liquid states\cite{ms}. Usually this transition occurs well 
inside the
supercooled region, where the liquid is already in a glassy state, 
and then it is related to polyamorphism\cite{ms,polya}. The
numerical modeling of these materials shows that interaction potentials can
be constructed for which there are transitions between different liquid
states. For instance, the potentials used to mimic the properties of water
generate two liquid phases in the supercooled region\cite{agua}. The main
characteristic of these materials (and of the interaction potentials used to 
model them)
which is responsible for their anomalous properties is the possibility of generating
open, low coordinated network structures at low pressure, while collapsing
to more compact structures at higher pressures\cite{ms}.

This key ingredient is captured by a class of model spherical potentials,
the so-called soft core Stell-Hemmer potentials\cite{sh}. They
include a strict hard core at some distance $r_{0}$ and a soft core at a
larger distance $r_{1}$. In this way, two typical distances between
particles occur, and there is a collapse from the largest to the smallest
distance on applying pressure. The properties of these systems give a
remarkably consistent description of most of the anomalous behavior of
tetrahedrally coordinated materials\cite{yo,sl,malescio}.

The possibility of liquid-liquid separation has been extended in the
last time by the finding that it can occur even at thermodynamic
equilibrium. Katayama {\it et al.}\cite{ff} 
report that phosphorus (which is a network
forming fluid) shows liquid-liquid coexistence at equilibrium, providing the
first experimental evidence of this phenomenon. In addition, Glosli and Ree\cite{carbon}
find numerical evidence of liquid-liquid coexistence at equilibrium in carbon.
This was the first direct numerical evidence of the phenomenon for a single
component system. The interparticle potential used to model carbon-carbon 
interaction is
non-isotropic, and incorporates information of the internal (electronic)
degrees of freedom of the atoms. It is shown that one of the two coexisting
liquids is mainly $sp$ coordinated, whereas the other shows a dominant $%
sp^{3}$ character.

On the basis of these results, it is natural to ask whether some kind of
soft-core potentials (with no internal degrees of freedom for the particles)
can produce an {\em equilibrium} liquid-liquid transition. 
We will show, by providing concrete examples, that soft core Stell-Hemmer potentials
are able to do that, then showing that the phenomenon may occur also for systems
that, in addition to being monodisperse, have particles which are spherical 
and `rigid',
i.e., in cases where internal degrees of freedom play no role. 
This is of basic interest since this model is much simpler, and then more 
transparent to analyze than real examples of material having liquid-liquid 
coexistence at equilibrium.
It also provides a benchmark for the study of
tetrahedrally coordinated
materials with liquid-liquid critical points in the supercooled region.
The relation between the existence of the
liquid-liquid critical point and the anomalous thermodynamic properties of
these materials can be studied here in a context of thermodynamic equilibrium.
On a different context, our results make plausible
the existence of liquid-liquid separation at equilibrium in 
particular cases of colloidal dispersions, which is of potential 
technological interest.

\section{The model}

We study a model of particles interacting through a potential which is the
sum of attractive and repulsive parts $V(r)=V_{A}(r)+V_{R}(r)$. The
repulsive part is given as 

\begin{eqnarray}
V_{R}(r)=\infty~~~~  {\rm for} ~~~~r<r_{0}\nonumber\\
V_{R}(r)=\varepsilon _{0}\left( r_1-r\right) /\left( r_{1}-r_{0}\right)~~~~
{\rm for} ~~~~r_{0}<r<r_{1}\nonumber\\
V_{R}(r)=0 ~~~~ {\rm for}~~~~ r>r_{1},
\end{eqnarray}
namely, a hard core at distance $r_0$, and a linear (repulsive) ramp between
$r_0$ and $r_1$.
The attractive part is
taken also to be linear in $r$, of the form 
\begin{eqnarray}
V_{A}(r)=-\gamma \left( r_2-r\right) /\left(
r_{2}-r_{0}\right) ~~~~ {\rm for}~~~~ r<r_{2}\nonumber\\
V_A(r)=0~~~~ {\rm for}~~~~ r>r_{2}
\end{eqnarray}
($r_{2}>r_{1}$, $\gamma >0$). 
The repulsive part of the potential favors one of the two
distances $r_{0}$ or $r_{1}$ between neighbor particles depending on the
value of external pressure and the attraction intensity $\gamma $.
Interparticle distance $r_{1}$($r_{0}$) is favored for pressures lower
(higher) than some crossover value $P_{0}.$ The crossover pressure
(for $\gamma =0$) is given roughly as $P_{0}=\varepsilon
_{0}/(v_{1}-v_{0})$, where $v_{0}$ and $v_{1}$ are the specific volumes of
the states with nearest particles at distances $\sim r_{0}$ or $\sim r_{1}$.
For $P$ near $P_{0}$ the system has an anomalously large
compressibility. The attractive part of the potential may turn this anomaly
into a first order transition through a van der Waals mechanism\cite{yo}.

We will show results for two-dimensional (2D) and three-dimensional (3D) 
systems. The parameters we use are:
$r_{1}=1.72r_{0}$, $r_{2}=4.8r_{0}$, and $\gamma=0.27\varepsilon_0$ for 2D and 
$r_{1}=1.72r_{0}$, $r_{2}=3.0r_{0}$, and $\gamma=0.31\varepsilon_0$ for 3D. 
Temperature will be measured in units of $\varepsilon_0/k_B$, and
pressure in units of $\varepsilon_0/r_0^2$ in 2D, and $\varepsilon_0/r_0^3$ in 3D.
Calculations were done by standard Monte Carlo techniques, in systems
with periodic boundary contitions. Some of the results to be presented where
done in the NVT ensemble, 
where the volume of the system is kept fixed at each simulated value,
and pressure is calculated through standard formulae. Other results correspond
to the NPT ensemble, where external pressure is fixed, and the volume of the system 
is taken as an additional variable during the simulation.

\section{Results}
\subsection{Two dimensions}
Fig. \ref{f1} shows the isotherms we get from the simulations for a system
of 200 particles in the 2D case, near the position in which we
find a liquid-liquid 
critical point for our parameters, 
namely $T_{LL}\simeq 0.10$, $P_{LL}\sim 0.07$, and
$v_{LL}\simeq 1.85 r_0^2$. We see in fact that when reducing 
$T$ below $\sim 0.1$, the isotherms
get a loop, typical of a first order transition. This transition separates two
liquid phases that we call (following the notation in water) low density liquid (LDL)
and high density liquid (HDL).
We have checked that the position and characteristics of this transition 
do not change when going to a system of 1000 particles.
As an example, we show in Fig. \ref{f1000}
the $T=0.105$ and $T=0.0875$ isotherms of systems 
of 200 and 1000 particles.
Isotherms are seen to coincide with the 200 particles case,
except within the coexistence region. Here the isotherm is more flat in the case of 
larger system. This is in fact what expected, since in an infinitely large system
isotherms are strictly flat in the coexistence region. Fig. \ref{snp}
shows snapshots of the
system at $T=0.0875$, within the LDL and HDL phases and in the coexistence region. 
The HDL (LDL) phase is characterized by a larger amount of particles 
at distance $r_{0}$ ($r_{1}$) from their neighbors. The snapshot at coexistence 
[Fig. \ref{snp}(b)] shows
indeed that different neighborhoods of high density and low density coexist 
in the system. The existence of a loop in the isotherms below the critical 
temperature, instead of a flat region, even when the system clearly separates in
different regions corresponding to the two different phases 
[Fig. \ref{snp}(b)] is due to the 
non-vanishing contribution to the total
energy of the surface energy between the two phases. 
This contribution only vanishes in infinite systems.

From the isotherms of Fig. \ref{f1} we see that the LDL
has a density anomaly, since there is a temperature range in which volume
increases as temperature is reduced, at constant pressure. This density
anomaly (which is reminiscent of the similar phenomenon in water) is known to 
occur\cite{yo}
in core-softened models even in the absence of an attractive part of the interparticle
potential, namely, when $U_A=0$.

We will now present the equilibrium $P$-$T$ 
phase diagram of the system, close to the
liquid-liquid critical point, to show clearly that the liquid-liquid 
coexistence we are
observing occurs in thermodynamic equilibrium, and it is not preempted by the
appearance of any other phase (gaseous or crystalline).

The $P$-$T$ phase diagram is presented in Fig. \ref{pt}, 
and the different borders between phases were obtained 
as explained in the following.
The position of the equilibrium first order line between LDL and
HDL phases in the $P$-$T$ phase diagram was determined  from the 
isotherms of Fig. \ref{f1} and from the values
of the enthalpy $h\equiv e+Pv$ (not shown) by thermodynamic integration, 
calculating differences in Gibbs free energy $G$ from the relation

\begin{equation}
\frac{G_2}{T_2}-\frac{G_1}{T_1}=\int_1^2 \frac{v}{T}dP-\frac{h}{T^2}dT
\label{g1}
\end{equation}
where 1 and 2 stand for two set of values $P_1$, $T_1$ and $P_2$, $T_2$
and the integration is through an arbitrary (reversible) path in th $P$-$T$ plane.
In this way we can compare $G$ on both sides of the first order line, performing
an integration surrounding the critical point, and the first order line can 
be determined with good precision. This procedure is more reliable than the standard
Maxwell construction, although the results we have obtained 
in this case are almost equivalent.

Concerning the crystalline structures, we know in fact 
that the present model shows a variety of different
crystalline structures, which are stable in different regions of the $P$-$T$
plane\cite{jagla1,jagla2}. 
When testing for the stability of these structures, the fact that
some of them take a very long time to appear spontaneously in simulations 
that start in the fluid 
phase is usually a problem, and then the equilibrium 
coexistence line between these phases and
the fluid has to be determined in some other way. 
We then applied the procedure already used in [11]
\begin{figure}
\narrowtext
\epsfxsize=3.3truein
\vbox{\hskip 0.05truein
\epsffile{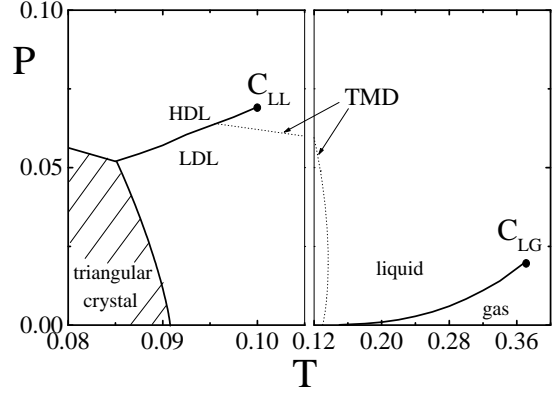}}
\caption{Pressure-temperature phase diagram of the two-dimensional
model showing the existence of the liquid-liquid and liquid-gas 
critical points. The temperature of maximum density (TMD) is indicated.}
\label{pt}
\end{figure}

consisting in the inclusion of 
an additional external potential, with an spatial periodicity
corresponding to the crystalline lattice to be studied. The existence of this
term (with a strength given by some parameter $W$) allows to construct a path
in the $P$-$T$-$W$ space that takes the system smoothly (i.e., without
crossing any phase transition) from the fluid to the 
crystalline phase, and the difference in free energy $\Delta G$
between the two phases at 
some fixed values $P=P_0$ and $T=T_0$ can
be calculated by a generalization of expression (\ref{g1}), namely

\begin{equation}
\Delta{(G/T)}|_{P_0,T_0}=\int_1^2 \frac{v}{T}dP+\frac{e_W}{TW}dW-\frac{h+e_W}{T^2}dT,
\label{g2}
\end{equation}
where $e_W$ is the potential energy per particle in the external potential. 
In order to move the system from the crystalline to the fluid phase, 
the integration consists typically in the following close path: increasing $W$
from zero to some large value; increasing $T$ from $T_0$ to a large value; 
decreasing $W$ down to 0; decreasing $T$ down to $T_0$.
In this way, the difference in free energy between the fluid phase and the 
possible crystalline phases was determined at fixed points in the $P$-$T$
phase diagram. Then $\Delta G$ can be  
calculated for the whole plane by thermodynamic 
integration of the pure phases.
Using this procedure we have verified that the only crystalline phase
that is thermodinamically stable within the $P$-$T$ sector shown in Fig. \ref{pt}
is the triangular structure, with a lattice parameter $\sim r_0$. Other 
crystalline structures appear only for lower temperatures of higher pressures.

Note in Fig. \ref{pt} that the melting of the crystal is anomalous 
in the region near liquid-liquid coexistence, i.e, both liquid phases 
are denser than the solid around this point, which is indicated by
the negative slope of the liquid-crystal coexistence line.
This is probably not a general rule, but is the case we found 
in all simulations we performed. In this figure the line of maximum density
is also indicated. Note that this line does not end exactly 
at the critical point, but somewhere on the first order line, within the
LDL region.

\begin{figure}
\narrowtext
\epsfxsize=3.3truein
\vbox{\hskip 0.05truein
\epsffile{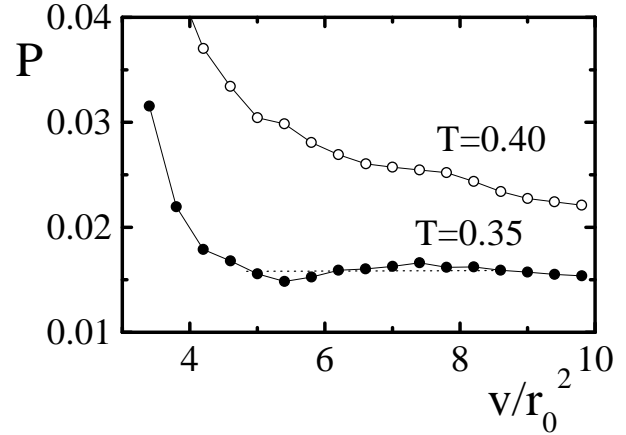}}
\caption{Isotherms near the liquid-gas transition for a two-dimensional
system of 500 particles.}
\label{lg}
\end{figure}

At very low pressure the condensed (liquid or
solid) phases are unstable against the formation of a gaseous phase, as in a 
standard fluid. The
condensed-gas coexistence line can be numerically determined in a reliable
way by the standard simulation methods we use only close to the (liquid-gas) 
critical point, where the metastability range
is small. Typical isotherms
around this point are shown in Fig. \ref{lg}, from which the location
of the liquid-gas critical point is estimated to be at $P_{LG}\simeq 0.02$, 
$T_{LG}\simeq 0.37$. 
Using the position of this critical point, we estimate the liquid-gas coexistence 
line by fitting to a van der Waals equation, chosen to give the correct values of
$T_{LG}$ and $P_{LG}$.
This is the line that is seen in the right part of Fig. \ref{pt} (note the change
in the horizontal scale in this sector).
At temperatures ($\sim 0.1$) where we observe 
liquid-liquid coexistence, the transition pressure to the gas phase
is of the order of $10^{-6}$.
Then we are sure that the liquid-liquid
transition occurs in a region in which liquid is thermodynamically stable
with respect to gas.

The crystal-liquid-liquid triple point we obtained turns out into a 
gas-liquid-liquid triple point
(still at equilibrium) by increasing the intensity of the attraction 
between particles. In
fact, this moves all coexistence lines to lower pressures compared 
to the  condensed-gas 
line, and at a
certain point the crystal-liquid-liquid triple point is forced to be in the  
metastable region with respect
to the gas phase. At this point, the gas-condensed line, together with the liquid-liquid line
determine a liquid-liquid-gas triple point. For instance, increasing $\gamma$ 
from 0.27 to 0.285 produces a liquid-liquid-gas triple point located at
$P\simeq 10^{-6}$, $T\simeq 0.10$. The liquid-liquid critical point is in this case
located at $P\simeq 0.03$, $T\simeq 0.12$.
At still higher values of the attraction the whole
liquid-liquid line moves into the metastable $P<0$ region, 
and the equilibrium liquid-liquid transition is
lost.

\subsection{Three dimensions}

\begin{figure}
\narrowtext
\epsfxsize=3.3truein
\vbox{\hskip 0.05truein
\epsffile{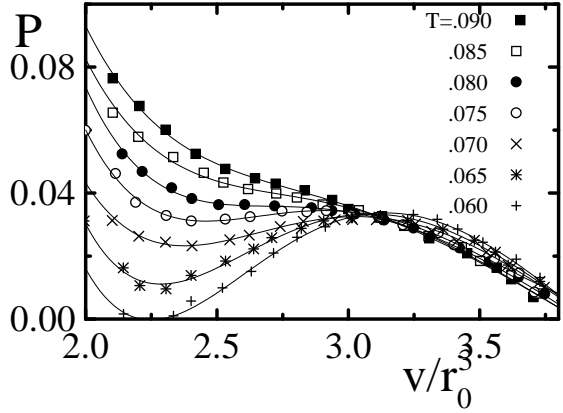}}
\caption{Isotherms of a 3D system of 300 particles 
near the liquid-liquid critical point
(located at $T\sim 0.08$, $P\sim 0.038$). Symbols are the simulated points,
lines are guides to the eye.}
\label{f3}
\end{figure}

For the 3D case, we show in Fig. \ref{f3} the isotherms obtained for a system of 300
particles. They again show the existence of the liquid-liquid critical point.
The curves are smooth for $T\gtrsim 0.08$, whereas there is a jump in
$v$ at the coexistence pressure for $T$ lower than this value. 
Also apparent from these curves is the existence (as in 2D) of density anomalies
for the low density liquid.

\begin{figure}
\narrowtext
\epsfxsize=3.3truein
\vbox{\hskip 0.05truein
\epsffile{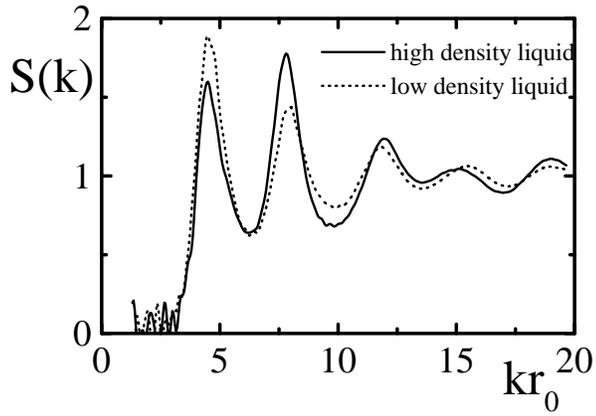}}
\caption{Structure factor of the two coexisting liquid phases of the 
3D system at $T=0.075$, $P\sim 0.032$.}
\label{f4}
\end{figure}

The structure factor $S(k)$ of the 3D liquid is shown for the two coexisting liquids at
$T=0.075$ in Fig. \ref{f4}. The two main peaks 
corresponding to interparticle distance $r_0$ and $r_1$ are indicated. There is an abrupt change 
in the relative weight of these two peaks at the liquid-liquid transition, corresponding to an abrupt
change in the mean distance between first neighbor particles. 

\begin{figure}
\narrowtext
\epsfxsize=3.3truein
\vbox{\hskip 0.05truein
\epsffile{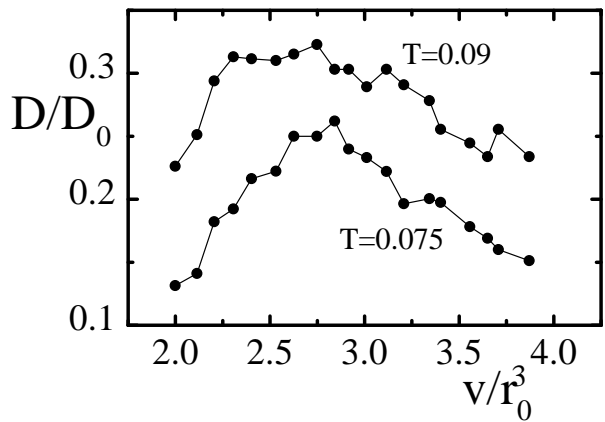}}
\caption{Relation between long and short time
diffusivity $D/D_0$ along the $T=0.075$ and $T=0.09$ isotherms, for the 3D system.}
\label{f5}
\end{figure}

For the simulations shown in Fig. \ref{f3} we did not 
find evidence of the appearance of any crystalline structure, 
suggesting  that the liquid-liquid coexistence 
occurs at equilibrium. To be sure, we should investigate the stability of the
different possible crystalline structures, as we did in the 2D case. For the
3D system however, this is a much harder work, as we actually do not know
all the crystalline structures for this case (see [12]). Then we take another
route, seeking for other indicators of stability of the
liquid phase, to rule out the possibility that we are 
observing a liquid-liquid transition in a metastable region. 
So we will look for the fulfillment of the Hansen-Verlet\cite{hv} and
Lowen\cite{lowen} criteria for crystallization. Although they stand only 
as approximate, and mainly heuristic, they
have been verified in a variety of cases.
For simple (i.e. Lennard Jones) systems, the Hansen-Verlet criterion states that freezing occurs when
the first peak of $S(k)$ reaches a (quasi-)universal value of $\sim 2.85$.
A straightforward generalization of this criterion to `anomalous' fluids tells that this relation should
be checked for all peaks in $S(k)$\cite{malescio}.
The results for $S(k)$ of Fig. \ref{f4} show that all peaks are
lower than this value, suggesting liquid stability. The Lowen 
criteria for freezing indicates that this occurs when the
relation between long and short times single particle diffusion 
coefficient drops below a value 
of $\sim 0.1$. This dynamical criterion is independent of the nature of the crystalline structure the
system freezes to. Although it has been proposed and verified in systems with 
Brownian dynamics, we have verified in simple cases (i.e. hard spheres) that it is
also fulfilled with the Monte Carlo dynamics, and so we apply it tentatively here.
The relation between long and short time values of the diffusivity
$D/D_0$ (calculated with the Monte Carlo dynamics)
is shown in Fig. \ref{f5} along the $T=0.075$ and $T=0.09$ isotherms. The relation $D/D_0$ is in 
the whole range well above the value of $0.1$ expected at freezing. Then all indicators coincide in telling that
the liquid-liquid coexistence we are observing occurs in truly thermodynamic equilibrium. 
From Fig. \ref{f5} we also see that the evolution
of $D$ on density shows a maximum which is well known to occur in tetrahedrally coordinates materials
\cite{diffexp}
and it is also typical of core-softened potentials\cite{younp}.

\section{Conclusions}

In conclusion, we have shown that a system of identical, spherical, and rigid particles is able to display
liquid-liquid coexistence at equilibrium (both in two and three dimensions). 
We showed explicitly this for an interaction potential that 
favors two different equilibrium distances between particles. 
The study of the present (or some related) model displaying liquid-liquid coexistence at equilibrium 
is likely to contribute to the understanding of the same phenomenon 
in cases where it occurs in the supercooled
region\cite{ms}, where it is much harder to study due to metastability against crystallization and long
relaxation times. Our findings also open the possibility of observing
liquid-liquid coexistence in monodisperse systems of rigid and spherical particles, i.e., colloidal systems, 
where interaction potentials with more than one equilibrium distance are easily obtained\cite{coloides}. 

\section{Acknowledgments}

This work was financially supported by Consejo Nacional de Investigaciones Cient\'{\i}ficas y
T\'ecnicas (CONICET), Argentina.
Partial support by Fundaci\'on Antorchas is also acknowledged.

\begin{figure}
\narrowtext
\epsfxsize=3.3truein
\vbox{\hskip 0.05truein
\epsffile{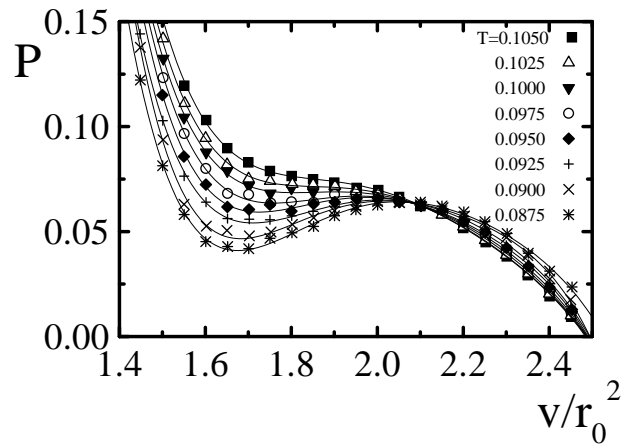}}
\caption{Isotherms from the simulation of a 2D system of 200 particles, at
temperatures as indicated (symbols are the simulated points, lines are
guides to the eye). There is a critical point at $T\simeq 0.10$, 
$P\simeq 0,07$, below which the isotherms get a typical van der Waals loop.
Note the expansion upon decreasing $T$ at constant $P$ (a density anomaly)
in the results for $v/r_0^2 {\protect \gtrsim} 2.1$.}
\label{f1}
\end{figure}

\begin{figure}
\narrowtext
\epsfxsize=3.3truein
\vbox{\hskip 0.05truein
\epsffile{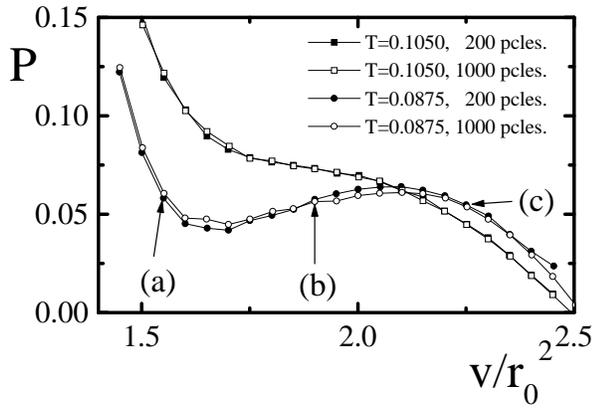}}
\caption{Comparison of isotherm obtained in simulations of systems with
200 and 1000 particles. They are coincident, except within the first order loop,
where the one corresponding to the larger system is more flat. This is the right
tendency since the isotherm at coexistence is absolutely flat in the limit
of systems of infinite size.}
\label{f1000}
\end{figure}

\begin{figure}
\narrowtext
\epsfxsize=3.3truein
\vbox{\hskip 0.05truein
\epsffile{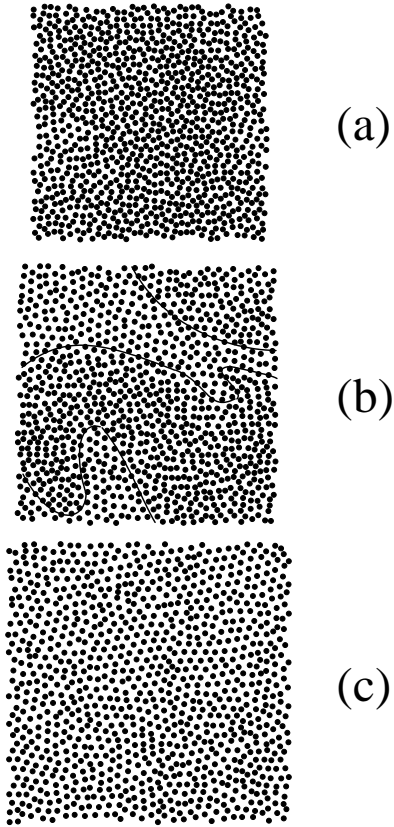}}
\caption{Snapshots of the system of 1000 particles at the points indicated in Fig.
\ref{f1000}. Black dots represent the strict hard core of the particles at $r_0$.
In (a) and (c) the system is almost completely in one of the
two liquid phases (HDL and LDL, respectively). In (b) there is a clear coexistence
of the two phases.} 
\label{snp}
\end{figure}

\end{document}